# Tailoring the Chirality of Microlaser with Topological Structured Chiral Droplets

Hamim Rivy [1] and Yu-Cheng Chen [1,2] *

[1] School of Electrical and Electronics Engineering, Nanyang Technological University, 50 Nanyang Avenue, 639798, Singapore

[2] School of Chemical and Biomedical Engineering, Nanyang Technological University, 62 Nanyang Drive, 637459, Singapore

*Correspondence E-mail: yucchen@ntu.edu.sg

**ABSTRACT**

The generation of chiral laser emission offers promising opportunities for modern photonic applications and the study of chiral light-mater interactions. Despite the great process made in recent years, the direct generation of chiral lasers with controllable chirality remains challenging in a microcavity. This study reports a strategy to control the emission chirality of whispering-gallery-mode organic microlasers with topological-structured chiral liquid crystal droplets. The findings suggest that the topological transformations in a microdroplet can induce different optical chirality strength and lasing polarization characteristics. In particular, the role of optical rotatory power was also investigated under linear and circular polarized excitation, where a vast lasing intensity difference between left/right circular polarized excitation was revealed. Theoretical analysis and simulation were carried out to support the significant findings. This study provides a facile tool for manipulating chiral laser emissions in microcavity, giving inspiration for topological-controlled photonics, chiral optical devices, and chiral molecular detection.

**Keywords**: chiral laser; circular polarization; whispering-gallery-mode; cholesteric liquid crystal; topological photonics

# INTRODUCTION

The interest in the generation of tunable circularly polarized (CP) laser sources has gained substantial attention in recent years. This is due to its immense applications in the potential development of some powerful and advanced photonic technologies such as chirality-based microscopy [1,2], chiral lasers [3], chirality-based sensors [4,5] and understanding of chirality-sensitive light-matter interaction [6,7]. To date, the development of circularly polarized laser radiation has been reported from chiral dyes [8,9], achiral dyes [10], periodic spatial modulation of achiral media [11,12], microsphere [13], metasurface [14,15], and cholesteric liquid-crystal (CLC) structures [16-19]. Among all, self-assembled liquid-crystal structures stand out as an ideal candidate for its controllability [20-25].

Cholesteric liquid crystals (CLC) have been extensively explored due to its distinctive optical properties and wide range of possible applications, especially in display technology [26], sensing [21,27], lasers [20,21,24,28], and photonic crystal fibers [29]. Another attractive feature of CLC is the capability to design microdroplets with the different topological states by adjusting its inner structure or external field [25,30-32]. Various studies have reported different topological configurations of liquid crystal droplets by adjusting its anchoring conditions or chiral dopant, which induces the rotation of the optical axis along a helical axis. A small change in chiral dopant leads to the change in chirality by changing the number of periodic refractive index variations, thus changes its topological state. While dye-doped CLC laser has been reported in many aspects, most of them were confined inside the photonic band gap region without a fixed chirality and topology [33-38]. Therefore, this work attempts to develop a chiral laser with tunable chirality and explore its lasing characteristics under topological diversity.

In this study, we report a distinct approach to control chiral lasing based on a WGM microresonator. We discovered that the lasing chirality and degree of chiral polarization could be manipulated by topological transformations in chiral droplets formed by CLC nanostructures. Four different chiral dopant concentrations were used to design four different topological structures from low chirality to high chirality (Fig. 1a). Both linear polarized and circular polarized excitation were explored in this work. Under linear polarized excitation, circularly polarized laser emission was observed at different wavelengths, and polarization was highly dependent on the chirality. As the pitch size becomes smaller, the optical rotatory power becomes larger and thus changes the polarization of laser emission. Under left and right circular polarized excitation (LCP/RCP), the droplets illustrated a definite dissymmetry factor, which is generally recognized as circular polarized emission in previous literature. Strikingly, laser emission was found to be nearly unpolarized regardless of the droplet chirality. Previous analysis on the generation CP laser emission from CLC structures mostly emphasized the study of dissymmetry factor and lacked a theoretical background. Nonetheless, our results reveal that the dissymmetry factor is insufficient to characterize the polarization of the output laser. The lasing polarization characteristics were elucidated based on light propagation theory, group velocity, and optical rotatory power of the topological chiral droplets. Theoretical analysis and simulation were carried out to support our findings. We believe this work may bring significant development in advanced photonic technologies, including chirality-based microscopy, tunable lasers, and chirality-based sensors.

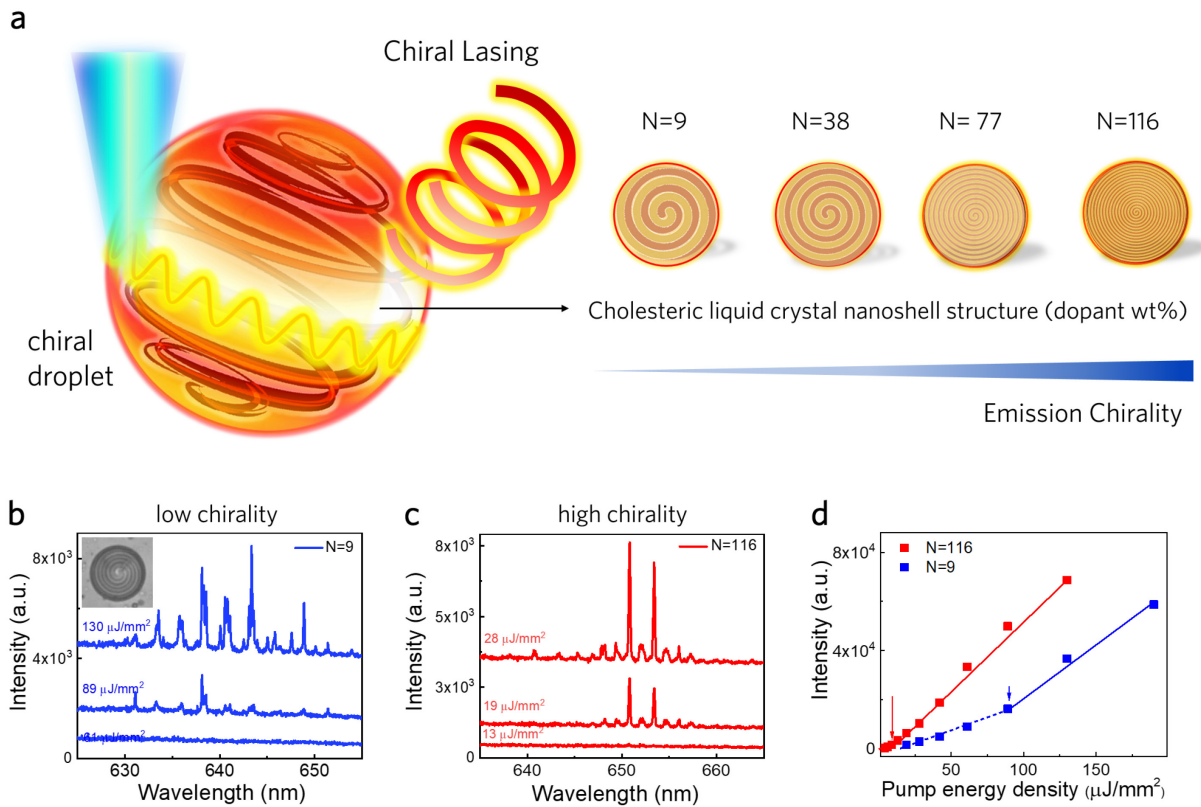

**Figure 1. (a)** The left schematic diagram shows a CLC droplet as WGM resonator, which can generate chiral laser emission. The right schematic represents CLC droplets with different topological diversity (N=9, N=38, N=77, and N=116), while the chirality was tuned by changing the chiral dopant concentration (1.2wt%, 5wt%, 10wt%, and 15wt %). The corresponding pitch sizes are 6.94 µm, 1.66 µm, 0.83 µm, and 0.55 µm. **(c-d)** Lasing spectra of CLC droplets at different pump energy density for **(b)** low chirality (N=9), and **(c)** high chirality (N=118) CLC droplets. **(d)** Spectrally integrated laser intensity as a function of pump energy density, extracted from Fig. 1(b) and 1(c). The color arrows indicate the respective lasing thresholds. Excitation = 488nm. All droplet diameters in this work are 32 µm (inset CCD of **b**).

# RESULTS

*Chiral lasing characteristics under linear polarized excitation*

At first, we started our investigation of laser emission by changing the chirality of the droplets under linear polarized excitation. We address the topological diversity issue by using CLC spherical WGM droplets with different chirality (low to high). The schematics are summarized in Fig. 1a, where we introduce the dimensionless parameter N=2d/*p*, which represents the number of pitch (*p*) turns of the director along the droplet diameter (d). The topological diversity in CLC nanostructures was controlled by varying the chiral dopant concentration to form different pitch sizes and N (N=9, N=38, N=77, and N=116). The CCD images of the CLC droplet with topological diversity are shown in Fig. S1. For instance, Figures 1b and 1c demonstrate the lasing spectra of CLC droplet under two different topological diversity, N=9 and N=116. For low chirality (N=9) droplets, the lasing peak was observed near 635 nm. On the other hand, lasing peaks were observed above 650 nm for high chirality droplets (N=116). The large red-shift exploited that wavelength tunability can be achieved by changing the chirality of the droplet. The effective refractive index of the droplets increased with increasing the chirality of the droplets, which conciliated the red-shift in the lasing wavelength. Taking a closer look at Figs. 1b and 1c, the free spectral range (FSR) for droplet (N=9) were measured to be 2.15 nm, which corresponds well with the WGM droplet diameter. However, as the topological diversity increases, the FSR becomes larger (~2.7 nm), which means the optical pathway is significantly decreased. This is likely the result of the light resonating in a spiral-like oscillation within the droplet where each path has more rotational power due to a higher amount of pitches.

Besides lasing spectra, Fig. 1d compared the spectrally integrated laser output as a function of pump energy density extracted from Fig. 1b and 1c. The corresponding lasing threshold was 89 µJ/mm$^2$, and 13 µJ/mm$^2$ for droplets with N=9, and N=116, respectively. As the chirality increases, higher lasing intensity and lower lasing threshold were found due to higher photon density of states (ρ) and light confinement. The enhancement of lasing intensity can be understood from the group velocity ($v_g$) of light inside CLC structures in two aspects. First, the group velocity ($v_g$) of the light decreases significantly near the photonic bandgap (PBG) region. Second, the group velocity ($v_g$) is inversely proportional to the photon density of states (ρ) based on $v_g = \frac{1}{\rho}$ [39, 40]. Hence the photon density of states (ρ) increases as it approaches the PBG wavelength region. In this respect, the PBG wavelength region becomes closer to the dye emission wavelength as the droplet chirality increased. A higher photon density of states will be achieved, therefore enhancing the intensity of laser emission and lower the lasing threshold. To validate this fact, the lasing intensity of topological droplets under different chirality (low to high) is compared in Fig. S2.

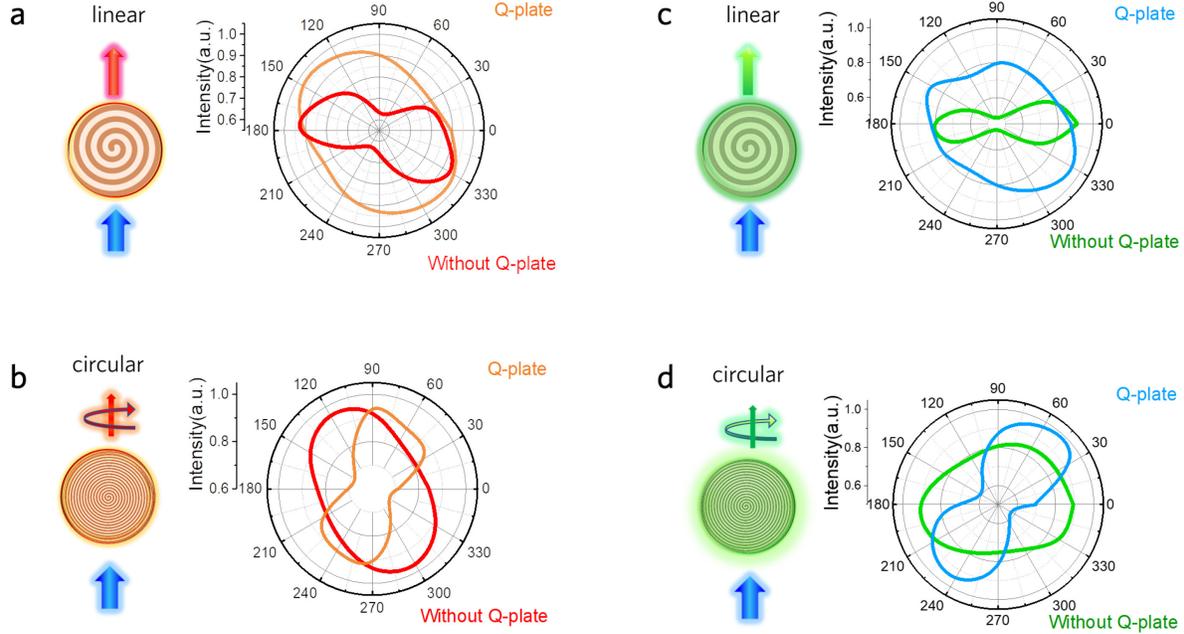

**Figure 2.** The polarization characteristics for topological chiral droplets under linear polarized excitation using **(a-b)** Nile red and **(c-d)** Coumarin-6 as the gain. Right-hand panel shows the output laser intensity as a function of the analyzer rotation angle θ (degree), without and with inserting a quarter (λ/4) waveplate in the beam path for **(a)** low chirality (N=9) and **(b)** high chirality (N=116) droplets doped with Nile Red. Excitation wavelength= 488nm. **(c-d)** Polarization characteristics for **(c)** low chirality (N=9) and **(d)** high chirality (N=116) droplets in Coumarin-6 doped spiral droplets. Excitation wavelength= 470nm. (Q-plate: quarter waveplate).

Next, we investigate how the topological transformations in chiral droplets may influence the polarization of WGM lasing emission. Figure 2 compares the polarization characteristics of the WGM laser emission generated from droplets with low and high chirality (doped with fluorescent dyes). Figures 2a and 2b show the spectrally integrated laser emission as a function of analyzer rotation angle for low chirality (N=9) and high (N=116) chirality droplets (Nile Red dye), under linear polarized excitation. The lasing intensity was collected twice by rotating the analyzer angle from 0º-360º with a 30º interval. Initially, WGM laser emission was collected directly without a quarter-wave plate. Later, laser emission was collected again subsequently by inserting a quarter-wave plate between the sample and the analyzer. A detailed experiment setup can be found in Fig. S3. For low chirality droplets (Fig. 2a), a strong modulation of the output laser intensity was observed without inserting the quarter waveplate. However, no intensity modulation was seen after passing the quarter waveplate, implying the output emission was linearly polarized (identical to the pump polarization) for low chirality droplets (N=9). To our surprise, an opposite scenario was observed for the droplet with high chirality (N=116) in Fig. 2b. Before inserting the quarter-wave plate, the output laser intensity was found to be independent of the rotation angle of the polarizer, which indicates that the generated laser was either circularly polarized or unpolarized. Subsequently, the lasing intensity was collected for the same droplet by inserting a quarter-wave plate. A huge intensity modulation was revealed by turning the laser emission into linearly polarized light. This result indicates that the generated red laser emission was circularly polarized or chiral after topological transformations to higher N or higher chirality. To demonstrate that such chiral lasing can be achieved at desirable wavelengths regardless of its topological structure, a similar experiment was conducted again by doping the droplets with a green emission dye (Coumarin 6). Figures 2c and 2d demonstrate that circular polarized laser emission can be adjusted at different wavelengths. The corresponding lasing spectra of such chiral lasing are supported in Fig. S4. Taking a closer look, the polarization of the generated laser remains similar to pump polarization for low chirality (N=9) droplets. In contrast, the polarization of the generated lasing becomes different from pump polarization for high chirality (N=116) droplets. This feature exhibits that circularly polarized laser can be generated at different the wavelength, which also overcomes one of the major problem (wavelength selectivity) introduced by synthesized chiral dyes.

Our findings motivated us to investigate the lasing chirality by manipulating the topological diversity of the droplet (N=9, N=38, N=77, and N=116). As the droplet chirality increases, the laser mode becomes more confined to the center region of the droplet (CCD images in Fig. 3a). In addition, the degree of circular polarization was measured, respectively. Herein DOCP was defined as $DOCP = 2.\left(\frac{I_{90}-I_0}{I_{90}+I_0}\right)$, where $I_{90}$ and $I_0$ are the integrated laser intensity after passing the quarter waveplate with the analyzer rotation angle 90° and 0° degree, respectively. Figure 3b presents the calculated degree of circular polarization (DOCP) as a function of the chirality. For low chirality droplets (N=9), DOCP was found to be 0.14, where DOCP =0.42 was found for high chirality droplets (N=116).

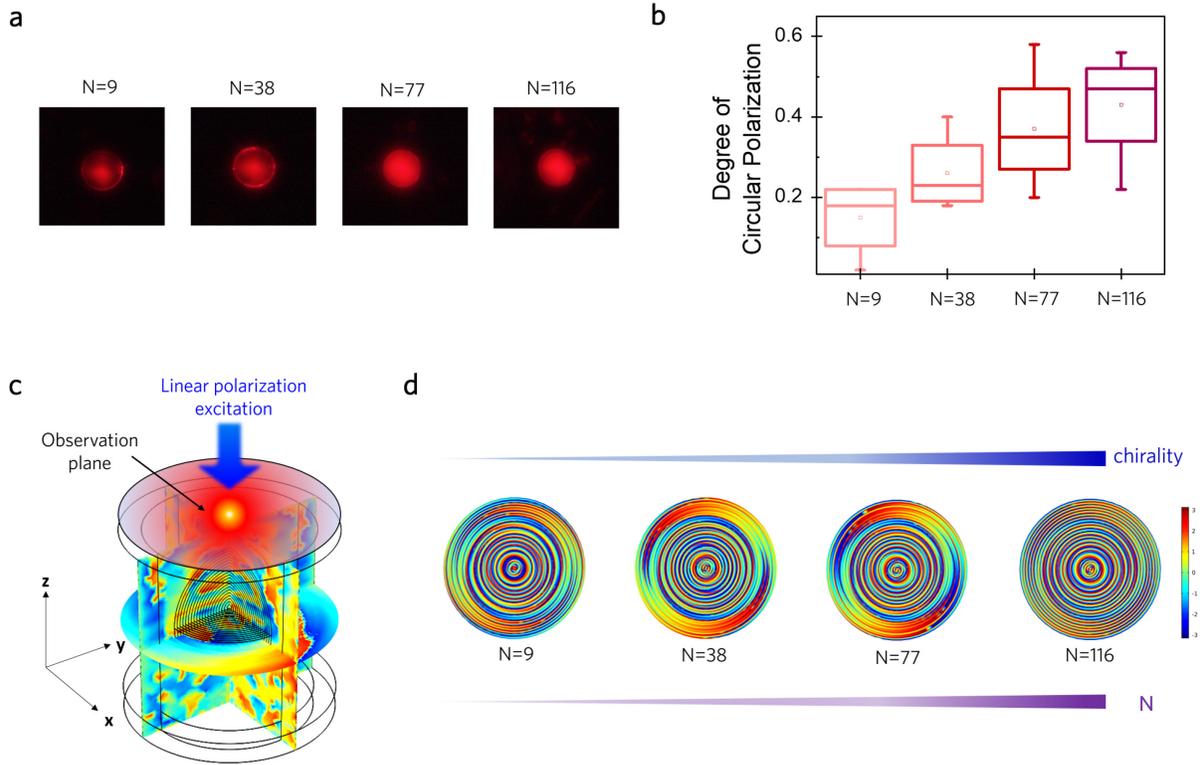

**Figure 3. (a)** Laser mode images captured by CCD. The images show the laser mode changed with increasing the chirality of CLC droplets. All droplet diameters = 32 μm. **(b)** The change of degree of circular polarization (DOCP) as a function of chirality of the CLC droplets from low chirality (N=9) to high chirality (N=116). All the data were collected over four randomly selected droplets for each chirality. **(c)** Schematic of the simulation setup and observation plane under linear polarized excitation. Here four CLC droplets were designed with increased chirality. **(d)** Phase distribution of reflected light in respective droplets under linear polarized excitation. CLC with four different chirality (low to high) were used where chirality was varied by changing the number of rings. The corresponding pitch sizes correspond to the pitch values used in this paper. Red color represents the phase of light, which is $180^0$ phase-shifted from the phase represented by blue color. Green (or near green) can be considered as $90^0$ phase-shifted from both red and blue. All figures present the light distribution at 620 nm emission wavelength.

Additionally, the phase distribution under linear polarized excitation for droplets with different chirality is shown in Fig. 3c. Different pitch sizes (*p*) were simulated corresponding to different chirality and N values (Fig. 3d). For linearly polarized excitation, the vortex-like phase distribution is observed and found to be 90º phase-shifted from each other, which indicates that the light reflected or transmitted from the CLC nanostructure experiences a 90º shifting. This phenomenon is similar to circularly polarized light, which can be considered as two combined electric fields that are 90º out of phase with each other. Furthermore, the phase distribution looks very similar for topological droplets with different chirality (N=9, N=38, N=77), but a little exception is observed for the droplet with extremely high chirality (N=116). The magnitude of the phase distribution was found stronger more at the center region compared to the outer shells (Fig. 3d). This behavior is observed mainly due to the decrease of optical pathway within the droplet, as the pitch size decreases. The CCD images in Fig. 3a shows that laser mode becomes more confined to the center region, as the pitch size decreases within the droplet, which mimics the behavior of simulation results. Although similar kind of phase distribution was observed for both low and high chirality droplets, only high chirality droplets show the possibility to generate high circularly polarized laser emission due to high optical rotatory power (see "Discussion" section).

*Lasing characteristics under circular polarized excitation*

In this section, we investigated the lasing and polarization characteristics of laser emission under left-hand circular polarized (LCP) and right-hand circular polarized (RCP) excitation. The schematic of Fig. 4a shows a topological droplet under circularly polarized excitation. Figure 4b clearly shows the feature of intensity as a function of the chirality of topological droplets (N=9, N=38, N=77, N=116) under LCP or RCP excitation at a fixed pump energy density. A huge intensity difference and signal-to-noise ratio were achieved between LCP and RCP. With the increment of droplet chirality, the spectrally integrated intensity increased for both LCP and RCP excitation. The integrated laser intensity was also found to be higher for LCP excitation compared to the RCP excitation. This phenomenon was mainly observed due to the type of chirality dopant (left-hand chiral dopant) within the CLC droplet. To obtain a fair comparison, data were collected over five randomly selected droplets with an identical diameter for each chirality. Figures 4c and 4d show the lasing spectra of droplets with the different topological transformation (low and high chirality) under LCP and RCP excitation. For both cases, laser emission was found to be red-shifted for the droplets with high chirality (N=118) compared to the droplets with low chirality (N=9), presenting a similar trend to the linear excitation in Figure 2. Given that similar wavelength and intensity tunability features were observed under linear and circular polarized excitation, vast differences were found in the polarization characteristics.

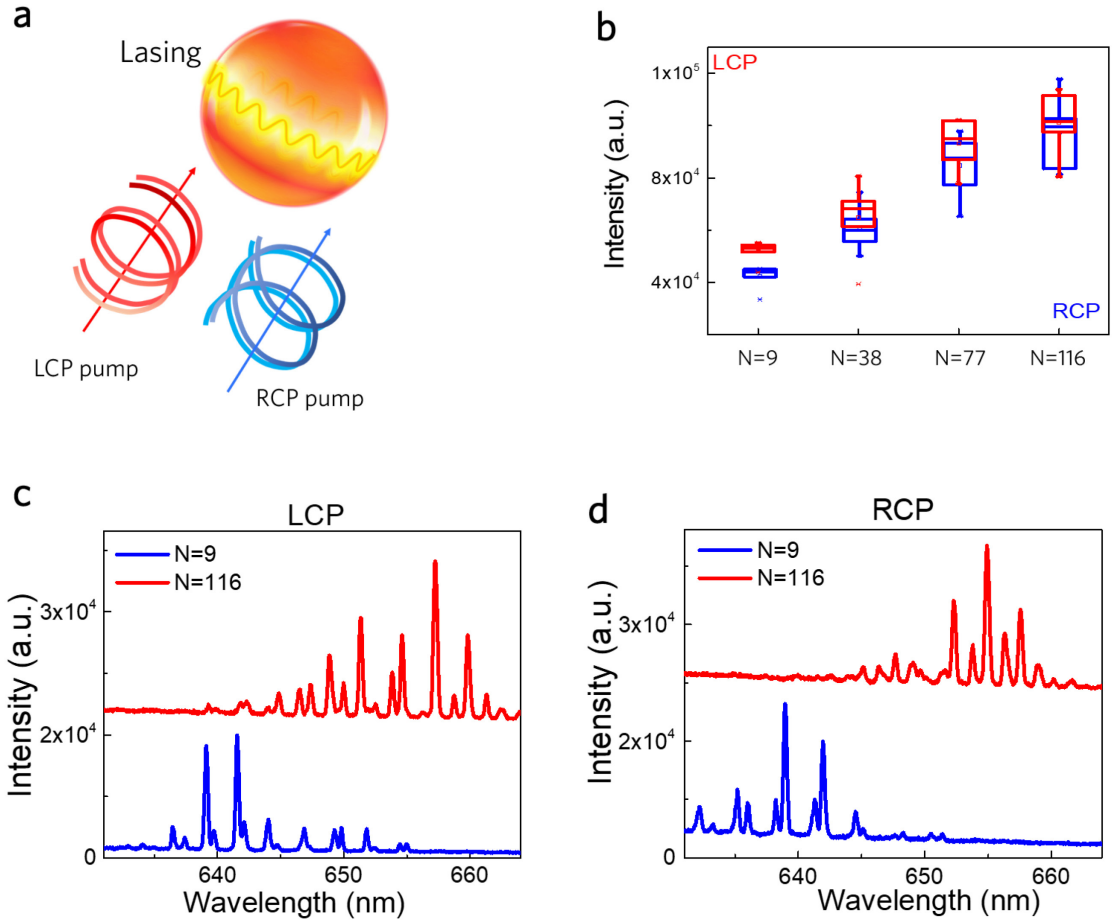

**Figure 4. (a)** Schematic of the topological chiral droplet as WGM resonator under left hand (LCP) and right hand (RCP) circular polarized excitation. **(b)** Spectrally integrated laser intensity with different topological diversity in chiral droplet (N=9, N=38, N=72, N=116) pumped under the same pump energy density (190 µJ/mm2). Data were collected over five random droplets (d=32µm) under LCP and RCP excitation. **(c-d)** Lasing spectra of droplets with two different chirality (N=9, and N=116) under (c) LCP and (d) RCP excitation, respectively. Excitation wavelength= 488nm. All droplet diameters = 32 µm. (Q-plate: quarter waveplate)

Figures 5a and 5b display the unique lasing spectra as a result of topological transformation in chiral droplets. At a specific pump energy density, a robust lasing signal could be achieved under LCP excitation, while no laser emission was obtained under RCP excitation for both low and high chirality droplets (N=9 and N=116). Figures 5c and 5d show the respective spectrally integrated laser intensity as a function of pump energy density. For left-hand excitation, a lower lasing threshold and higher intensity were observed in both cases compared to be right-hand excitation. This indicates the existence of the dissymmetry factor, which is basically the differential emission intensity between left (LCP) and right circular polarized (RCP) light. In this work, dissymmetry factor is defined as $|g_{dis}| = \frac{2(I_{L-slope} - I_{R-slope})}{I_{L-slope} + I_{R-slope}}$. For low chirality droplet, dissymmetry factor was found to be $|g_{dis}| = 0.19$, where dissymmetry factor was $|g_{dis}| = 0.43$ for high chirality droplet. The magnitude of the dissymmetry factor is the most common measure to show the degree of circular polarization. As low and high chirality droplets showed an absolute magnitude of dissymmetry factor (0.19 and 0.43, respectively), the emitted laser should have a certain degree of circular polarization. While most articles considered this phenomenon as circularly polarized emission [16-19], our study implied an utterly different observation. As shown in Figs. 5e and 5f, the generated laser emission was found to be nearly unpolarized for high chirality droplets (Fig. 5f), as it does not show any intensity modulation with or without adding any quarter waveplate. Contrarily, Fig. 5e shows a small degree of circular polarization (similar to the pump polarization) for the low chirality, although the dissymmetry factor was lower.

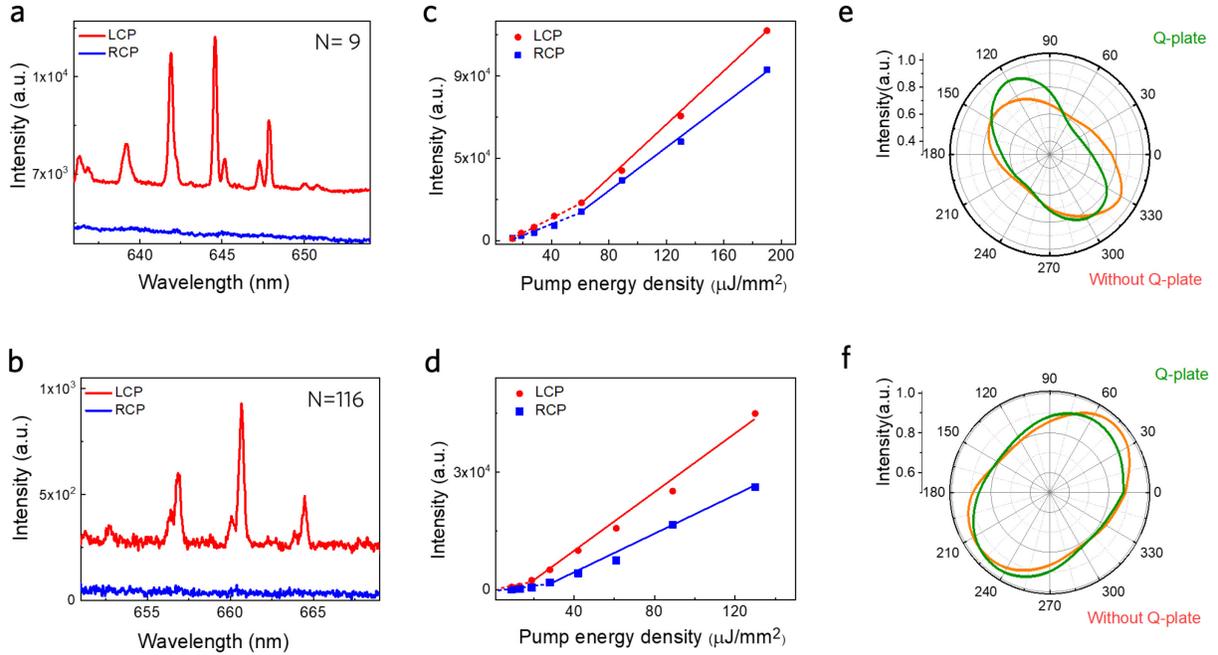

**Figure 5. (a-b)** Laser emission spectra of the chiral droplet at relatively low pump energy density under left hand (LCP) and right hand (RCP) circular polarized excitation for **(a)** low chirality (N=9) and **(b)** high chirality (N=116) droplets. **(c-d)** Spectrally integrated laser output as a function of the pump energy density under left-hand and right-hand circular polarized light for **(c)** low chirality (N=9) and **(d)** high chirality (N=116) droplets. Solid curves are the linear fit for the above lasing threshold, where doted curves are fitting below the threshold. **(e-f)** Spectrally integrated output laser intensity as a function of the analyzer rotation angle for **(e)** low chirality (N=9) **(f)** high chirality (N=116) droplets under fixed pump energy density.

## DISCUSSION

The polarization characteristics under circular and linear polarized excitation illustrated that lasing generated from the low chirality droplets possess identical polarization as the pump laser. In contrast, droplets with high chirality could strongly alter the polarization state of the pump laser. The polarization characteristics can be explained by the optical rotatory power of the CLC. The inset of Fig. 1b shows that our structure is a radial spherical structure which has a defect line in the center. The defect line is due to the chiral dopant induced twist in the liquid crystal, which leads to a helical organization of the molecules. A double spiral disclination line forms at the interface of the droplets. It reflects the handedness and spiral distribution of the director field in the bulk of the droplet [41]. According to previous literature, [42] we can assume the droplets have a helical structure along the circumference of the spherical structure. As our cavity generates WGM laser, we can consider light propagates along the boundary (the helical axis) of the spherical structure. From the theory of light propagation along the helical axis of CLC, the optical rotatory power can be written as [43, 44]

$$\rho = \varepsilon_a^2 k^4 / 8q \, (q^2 - k_o^2)$$

where $\rho$= optical rotatory power, $\varepsilon_a$ =extraordinary permittivity, k=wave vector, $k_0$= wave vector at the direction of the ordinary refractive index, and $q=2\pi/p$. p= pitch of the CLC structure. From the equation, it can be observed that the large pitch compared to the wavelength leads to a minimal value of q. It results in a deficient optical rotatory power in CLC structure along the helical axis. On the other hand, when the pitch (p) is comparable to the wavelength, the equation results in a very high optical rotatory power. This phenomenon agrees well with our experiment result. Within our experiment, we have seen that the droplets with very low chirality (p=6.94μm) or low topological transformations have very little light modulation strength. As a result, the generated laser polarization was mainly dominated by its initial pump polarization. Contrarily, high chirality droplets (high topological transformation) with higher rotatory power change the polarization states of the output laser emission. Therefore, circularly polarized laser emission was achieved.

# CONCLUSION

In this study, we explored the possibility to generate and manipulate the laser emission chirality by using CLC droplets with different topological transformations. Both experimental and theoretical studies revealed how the topology and resultant chirality of the droplets affects the output laser polarization states under linear and circular polarized excitation. Our findings show that the topological transformation inside a microdroplet can significantly induce different chirality strength or lasing polarization. As such, droplets with high chirality possessed high optical rotatory power, which could change the polarization state of the output laser. Therefore, tunable circularly polarized laser emission was achieved from the droplets with high chirality, under linear polarized excitation. Contrarily, laser emission was found to be nearly unpolarized under circular polarized excitation. It is envisaged that this study might shed light on the development of powerful photonic devices by tuning the topological state in a microcavity. Potential applications for such chiral droplet laser include chirality-based imaging and chirality-based sensors.

# METHODS & MATERIALS

Cholesteric liquid crystal (CLC) droplets with different topological transformations were prepared from high chirality to low chirality. Four different chiral dopant concentrations were used (1.2wt%, 5wt%, 10wt%, and 15wt %) to achieve droplets with different topological diversity (N=9, N=38, N=77, and N=116). S-811 (4-[1-methylheptyloxy] carbonylphenyl-4-(hexyloxy) benzoate) was used as chiral dopant. Different concentrations of chiral dopant (S-811) were mixed with nematic liquid crystals (5CB, from Tokyo Chemicals) and 0.2 wt% fluorescence dyes. Nile Red or Coumarin-6 were used as fluorescence dyes. Later, the mixture was vortexed for 5 minutes and immersed in an ultrasonic bath for 10 minutes to form a homogenous CLC solution. To from the droplet, the CLC mixture was added in a surfactant solution at a 1:25 ratio. Polyvinyl alcohol (PVA) (1wt%) in DI water was chosen as the surfactant. It facilitates the WGM mode in CLC droplet, as its refractive index (RI) is lower than the effective RI of 5CB ($n_{eff}$=1.625 at 650 nm wavelength). Finally, the solution was vortexed for 45 seconds, and droplets were prepared. A droplet size of 32μm on average was selected for all experiments for an optimized q-factor [45]. Most of the droplets were radial spherical structures with s = 2 disclination lines, shown in the inset of Fig. 1b. All chemicals were purchased from Sigma-Aldrich unless specified above.

The experimental setup is shown in Fig. S3. An inverted microscopic system (Nikon Ti2) with 20X 0.4 NA objective was used to excite the microcavity and collect of laser emission. A pulsed ns-laser (EKSPLA PS8001DR) integrated with an optical parametric oscillator (repetition rate: 50 Hz; pulse duration: 5 ns) was used to achieve optical pump. To match the dye absorption wavelength, the pump was tuned at 488 nm and 470 nm for Nile Red and Coumarin-6, respectively. A circular polarized laser pump was achieved by adding circular polarizer (Right hand and Left hand) in the beam path. Circular polarizers were bought from Thorlabs (488 nm). Laser emission was collected twice to check the polarization characteristics of laser. Firstly, after passing through the analyzer. Secondly, after passing through the quarter waveplate and the analyzer. Wavelength specific red (633nm) and green (543 nm) zero-order quarter waveplates (Thorlabs) were used. The collected light was sent into an imaging spectrometer (Andor Kymera 328i and Newton 970 EMCCD).

## ACKNOWLEDGMENT

We would like to thank the lab support from Centre of Bio-Devices and Bioinformatics and Internal Grant NAP SUG - M4082308.040 from NTU.

**The Authors Declare No Conflict of Interests.**